\documentclass[10pt,aps,prl,reprint,superscriptaddress,floatfix]{revtex4-2}
\usepackage{amsmath,amssymb,bm}
\usepackage{graphicx}
\usepackage[hidelinks]{hyperref}
\begin{document}

\title{\texorpdfstring{Giant Thermal-Conductivity Enhancement from Chiral-Phonon Pseudo-Angular Momentum Conservation}{Giant Thermal-Conductivity Enhancement from Chiral-Phonon Pseudo-Angular Momentum Conservation}}

\author{Tingting Wang}
\affiliation{Ministry of Education Key Laboratory of NSLSCS, Phonon Engineering Research Center of Jiangsu Province, Center for Quantum Transport and Thermal Energy Science, Institute of Physics Frontiers and Interdisciplinary Sciences, School of Physics and Technology, Nanjing Normal University, Nanjing 210023, China}
\affiliation{Department of Physics, National University of Singapore, Singapore 117542, Singapore}

\author{Dabao Zha}
\affiliation{Ministry of Education Key Laboratory of NSLSCS, Phonon Engineering Research Center of Jiangsu Province, Center for Quantum Transport and Thermal Energy Science, Institute of Physics Frontiers and Interdisciplinary Sciences, School of Physics and Technology, Nanjing Normal University, Nanjing 210023, China}

\author{Hao Chen}
\email{phych@njupt.edu.cn}
\affiliation{School of Science, Nanjing University of Posts and Telecommunications, Nanjing 210023, China}

\author{Jiangbin Gong}
\email{phygj@nus.edu.sg}
\affiliation{Department of Physics, National University of Singapore, Singapore 117542, Singapore}

\author{Lifa Zhang}
\email{phyzlf@njnu.edu.cn}
\affiliation{Ministry of Education Key Laboratory of NSLSCS, Phonon Engineering Research Center of Jiangsu Province, Center for Quantum Transport and Thermal Energy Science, Institute of Physics Frontiers and Interdisciplinary Sciences, School of Physics and Technology, Nanjing Normal University, Nanjing 210023, China}

\begin{abstract}
Pseudo-angular momentum (PAM) underlies optical selection rules for chiral phonons, but whether it also constrains thermally populated finite-wave-number phonon-phonon scattering has remained unresolved. We show that rotational or screw eigenphase conservation imposes a PAM residue rule on cubic anharmonic vertices, revealing a hidden selection rule for heat transport. In screw-symmetric helical Te, an exact platform, implementing this rule as a projector in first-principles Boltzmann transport leaves spectra and force constants unchanged but removes roughly two thirds of kinematically allowed triplets, suppresses resistive Umklapp relaxation, and enhances lattice thermal conductivity by a factor of \(5.30\) at 300~K, remaining above fivefold up to 400~K. A bulk chiral-crystal benchmark further shows that explicit eigenphase organization can increase the calculated lattice thermal conductivity by about \(24\%\), comparable to the reported first-principles underestimation of experiment. These results establish PAM conservation as an anharmonic selection principle for chiral-phonon heat transport and as a fundamental principle to guide the prediction and control of thermal conductivity in chiral crystals and nanoscale phononics.
\end{abstract}

\maketitle

\textit{Introduction.--} Chiral phonons endow lattice vibrations with handedness and angular-momentum-like structure~\cite{zhang2015chiral,zhu2018observation,wang2024chiral,juraschek2025chiral}, including phonon angular momentum~\cite{zhang2014angular}. Their handed lattice motion couples to magnetic, spin, valley, and topological responses~\cite{juraschek2019orbital,kim2023chiral,wang2022chiral,pan2023vibrational,luo2023large}. In crystals with rotational or screw symmetry, the symmetry quantum number that organizes this handed motion is pseudo-angular momentum (PAM), the eigenphase of a phonon mode under a crystal operation~\cite{streib2021difference,zhang2022chiral}. PAM underlies the optical selection rules that make chiral phonons addressable by circularly polarized probes~\cite{zhu2018observation,ishito2023chiral,spirito2024lattice,oishi2024selective}. The central question for heat flow is whether this same eigenphase quantum number also acts as a hidden conservation law for thermally populated finite-wave-number, or finite-$q$, phonon-phonon scattering.

Conventional three-phonon transport calculations enforce energy conservation and translational crystal-momentum conservation, but a nonzero cubic vertex in a rotation- or screw-symmetric lattice must also balance the eigenphases of the participating phonons. This missing layer is especially relevant for chiral crystals and quasi-one-dimensional screw lattices. Chiral phonons in Te are well established~\cite{pine1971raman,ishito2023chiral,spirito2024lattice}, one-dimensional or confined Te chains realize a $3_1$ screw lattice~\cite{medeiros2017single,poborchii2023optical,liu2018allotropes}, and single-crystalline Te nanowires with chirality-dependent responses have been demonstrated experimentally~\cite{Calavalle2022TeNanowire}. In $\alpha$-quartz, first-principles Boltzmann calculations reproduce the anisotropy of lattice thermal conductivity but underestimate measured values despite neglecting defect scattering~\cite{Mizokami2018SiO2ThermalConductivity}. In trigonal Te, electronic and sample-dependent effects can obscure the intrinsic lattice contribution~\cite{peng2015anisotropic,Ho1972ThermalConductivityElements}. These comparisons point to a symmetry-resolved constraint that is usually left implicit in thermal-transport workflows: rotational or screw eigenphase conservation can supplement spectra, group velocities, force constants, and energy--momentum phase space as a selection rule for anharmonic relaxation.

In this Letter, we make this hidden conservation law explicit and showcase its important physical implications for anharmonic heat transport. We use a dynamically stable one-dimensional $3_1$ helical Te chain as a material-relevant exact proof platform: because the full Brillouin zone is screw invariant, every phonon mode carries a screw-PAM label and the cubic-vertex nonzero condition can be imposed without approximate mode labeling. We derive the residue rule from screw-eigenphase conservation, implement it as a projector in first-principles Boltzmann transport, and compare the PAM-conserving channel set with a PAM-relaxed kinematic baseline. This construction lets us trace the mechanism from symmetry-allowed triplets to Umklapp relaxation, lifetimes, mean free paths, thermal conductivity, and PAM-velocity-locked directional propagation; a compact bulk $\alpha$-quartz benchmark tests whether the same eigenphase-organized scattering picture has material-scale relevance beyond the exact chain.

\textit{Symmetry principle and transport calculation.--} A chiral phonon may carry local mechanical angular momentum, but the conserved quantity that controls an anharmonic selection rule is the eigenphase of a crystal operation. For the $3_1$ helical chain, the relevant operation is
\begin{equation}
 g=\{C_3|c/3\},
\end{equation}
where $c$ is the chain period. Here and below, $q$ is the scalar Bloch wave number along the chain axis, with reciprocal vector $G_z=m\,2\pi/c$. Since the present system is one-dimensional, the full Brillouin zone is the screw-invariant line. A phonon mode can therefore be chosen as an eigenstate of $g$ with
\begin{equation}
\chi_j(q)=\exp\left[-i\left(\frac{qc}{3}+\frac{2\pi}{3}l_{s,j}\right)\right],
\label{eq:chi_prl}
\end{equation}
where $l_{s,j}=0,\pm1$ is the spin-like intracell PAM. The first term is the orbital/intercell phase generated by the fractional translation, while the second is the intracell PAM phase. This separation is crucial: the transport rule involves $l_s$, but the screw-translation phase is retained through the Umklapp index.

For a three-phonon absorption process $1+2\rightarrow3$, let \(V^{(+)}_{123}\) denote the corresponding cubic anharmonic vertex in the phonon eigenmode basis. Screw invariance gives
\begin{equation}
V^{(+)}_{123}=\chi_1(q_1)\chi_2(q_2)\chi^*_3(q_3)V^{(+)}_{123}.
\end{equation}
A nonzero vertex therefore requires
\begin{equation}
\chi_1(q_1)\chi_2(q_2)\chi^*_3(q_3)=1.
\end{equation}
Combining this condition with crystal-momentum conservation, $q_1+q_2-q_3=G_z=m\,2\pi/c$, gives
\begin{equation}
l_{s1}+l_{s2}-l_{s3}+m\equiv0\pmod 3.
\label{eq:pam_abs_prl}
\end{equation}
For emission, $1\rightarrow2+3$, the corresponding rule is
\begin{equation}
l_{s1}-l_{s2}-l_{s3}+m\equiv0\pmod 3.
\label{eq:pam_em_prl}
\end{equation}
Thus the PAM residue rule is the representation-level nonzero condition for the cubic anharmonic vertex. The fractional-translation phase is retained through the ordinary Umklapp index \(m\), while \(l_s\) carries the intracell eigenphase balance. The full derivation from the cubic Hamiltonian is given in Sec.~S1 of the Supplemental Material~\cite{SM}, whereas Sec.~S2 gives the general $C_n$ rotation/screw form and its extension to higher-order anharmonic processes.

\begin{figure}[!tbp]
    \centering
    \includegraphics[width=\columnwidth]{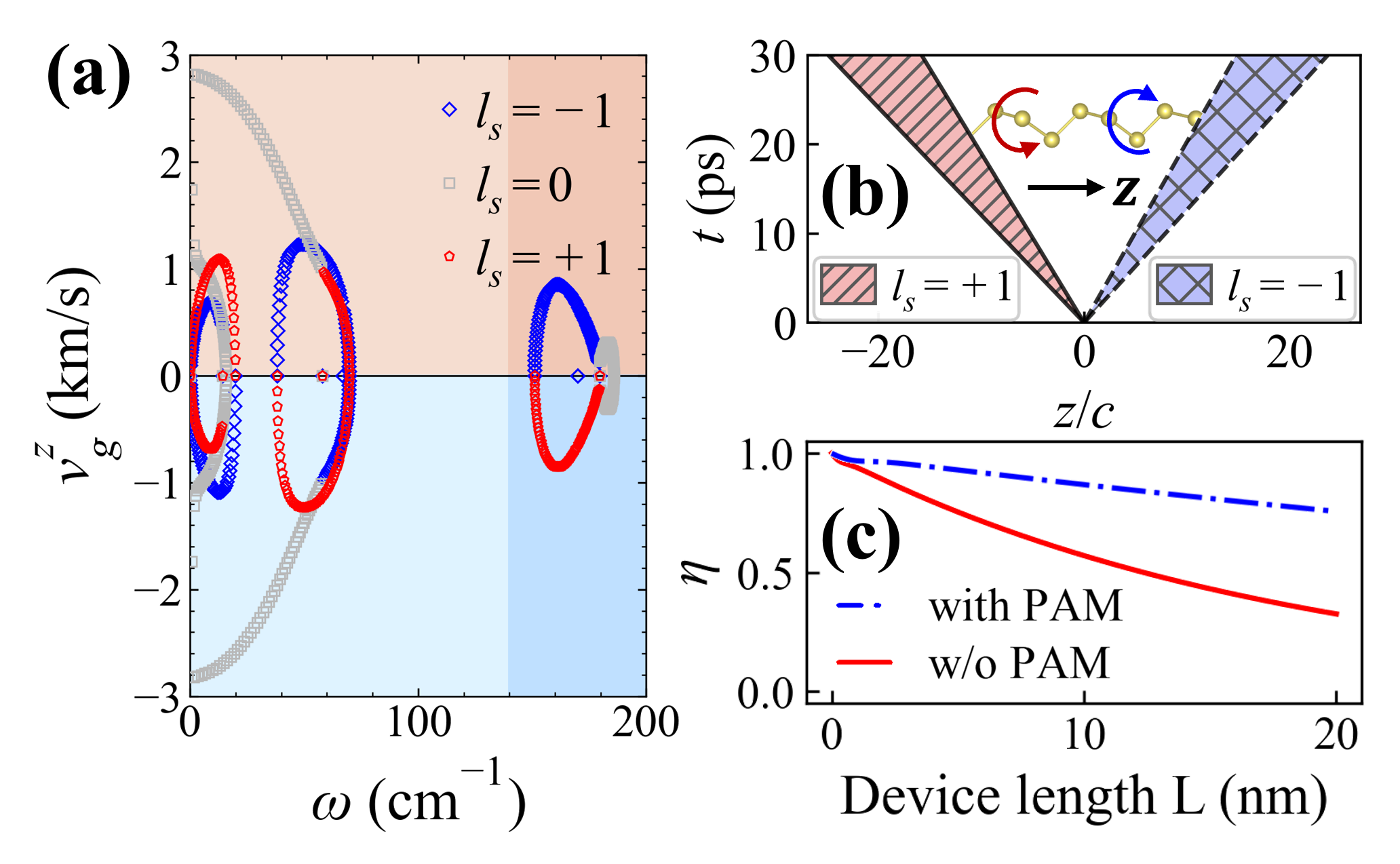}
    \caption{Rectification estimate from PAM-velocity locking and PAM-conserving lifetimes. (a) Group velocity $v_g^z$ versus frequency, labeled by $l_s=-1,0,+1$; shaded regions mark optical windows where the sign of $v_g^z$ is locked to the PAM label. (b) Ideal ballistic propagation fan for the two opposite optical chiralities, showing the separation of $l_s=+1$ and $l_s=-1$ wave packets in space and time. (c) Rectification estimate $\eta$ as a function of device length, comparing the PAM-conserving physical calculation with the diagnostic baseline labeled ``w/o PAM'' in the figure.}
    \label{fig:5}
\end{figure}

We make Eqs.~(\ref{eq:pam_abs_prl}) and (\ref{eq:pam_em_prl}) explicit in first-principles three-phonon Boltzmann transport calculations. Harmonic and third-order force constants are obtained from finite displacements, with the low-dimensional symmetry refinement and numerical settings summarized in Sec.~S3 of the Supplemental Material~\cite{SM}; the first-principles and transport tools are described in Refs.~\cite{kresse1996efficient,gonze1997dynamical,phonopy-phono3py-JPCM,eriksson2019hiphive,li2014shengbte}. The mode-resolved PAM $l_s$ is extracted from the screw eigenphase of the phonon eigenvectors, including subspace diagonalization for near-degenerate modes, as described in Sec.~S3. In the transport workflow, we apply the residue rule as a projector after the usual energy- and crystal-momentum checks. The PAM-conserving channel set and the PAM-relaxed baseline therefore share identical spectra, third-order force constants, occupations, and kinematic phase space; they differ only by the screw-eigenphase nonzero condition. We refer to the latter diagnostic baseline as ``w/o PAM'' in the figures. To connect the projected scattering channels to transport, we use three diagnostics: $W_3(\omega)$ measures the temperature-weighted three-phonon scattering phase space, $\kappa_c(\omega)$ is the cumulative lattice thermal conductivity below frequency $\omega$, and $\Lambda_\lambda=|v^z_\lambda|\tau^{\rm eff}_\lambda$ is the mode mean free path. For the directional-propagation estimate in Fig.~\ref{fig:5}, the same mode lifetimes enter a survival factor $\exp[-L/\Lambda_\lambda]$, and $\eta$ denotes the resulting right-left propagation imbalance. Further computational details are given in the Supplemental Material.

\begin{figure}[!tbp]
    \centering
    \includegraphics[width=0.82\columnwidth]{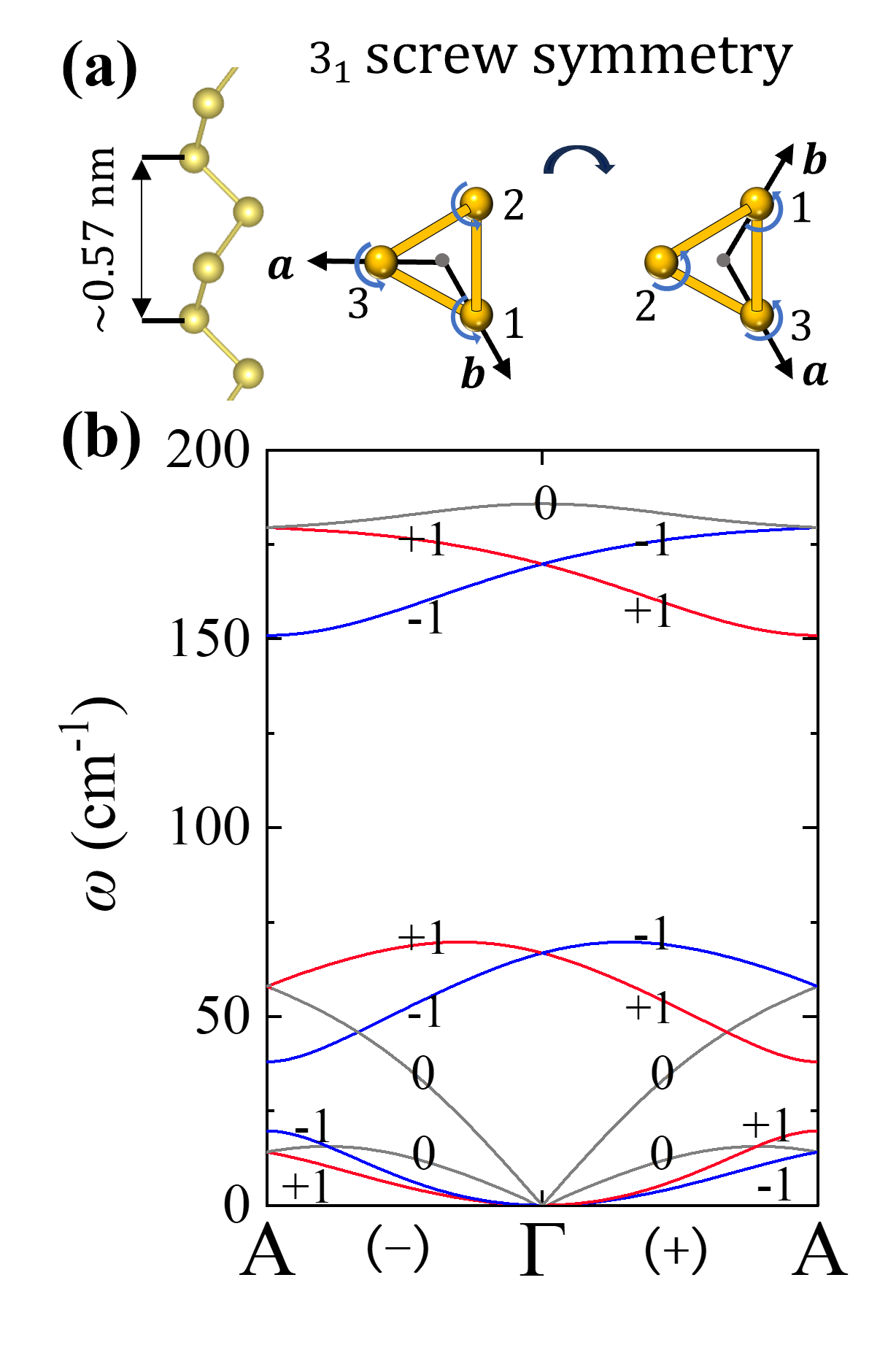}
    \caption{Intrinsic chiral phonons in the one-dimensional $3_1$ helical Te chain. (a) Right-handed helical structure and transverse triangular Te motifs related by the screw operation $\{C_3|c/3\}$. Curved arrows indicate the handed circular motion associated with chiral phonon sectors. (b) Phonon dispersion along $\mathrm{A}$--$\Gamma$--$\mathrm{A}$, labeled by the spin-like PAM quantum number $l_s=0,\pm1$. The absence of imaginary branches confirms dynamical stability.}
    \label{fig:1}
\end{figure}

\textit{Results and discussion.--} Figure~\ref{fig:1} establishes the structural and single-phonon basis. The relaxed Te chain is a genuine $3_1$ screw lattice, not a geometrically twisted version of an achiral chain. Its phonon spectrum has no imaginary branches, confirming dynamical stability. The optical branches are cleanly separated into $l_s=+1$, $l_s=-1$, and $l_s=0$ sectors, corresponding to opposite circular-polarization sectors and a symmetry-neutral sector. Near $\Gamma$, the acoustic modes are dominated by long-wavelength translations, but their symmetry labels still connect continuously to the same screw-eigenphase classification. This is important because the heat carriers are not necessarily the most visibly chiral optical modes. Rather, optical chiral modes can act as symmetry-selective scattering partners for low-frequency heat-carrying phonons.

\begin{figure}[!tbp]
    \centering
    \includegraphics[width=\columnwidth]{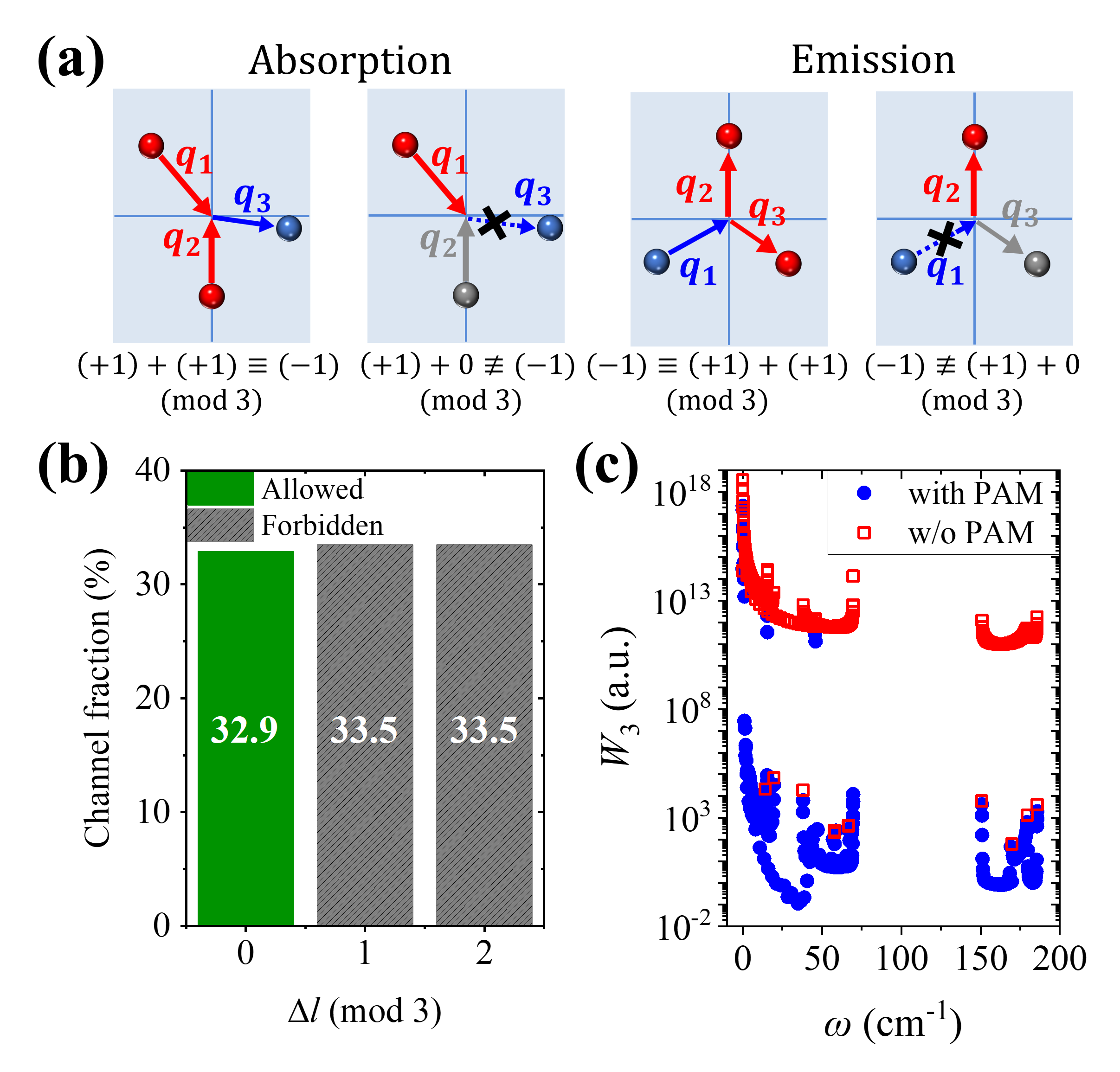}
    \caption{PAM selection rule and phase-space reduction. (a) Representative allowed and forbidden three-phonon absorption and emission processes. The equations below each schematic show the corresponding modular PAM-balance condition for the normal-like examples shown. (b) Distribution of the reduced PAM residue $\Delta l\equiv\Delta_{\rm PAM}$ (mod 3), showing that only the $\Delta l=0$ class is compatible with a nonzero screw-symmetric cubic vertex. (c) Temperature-weighted three-phonon phase space $W_3(\omega)$ at 400~K, comparing the PAM-conserving physical channel set with the diagnostic PAM-relaxed baseline (``w/o PAM'').}
    \label{fig:2}
\end{figure}

Figure~\ref{fig:2}(a) illustrates how this mode label becomes a scattering rule. The representative absorption and emission processes are chosen so that the ordinary momentum construction looks similar for allowed and forbidden cases; what changes is the PAM residue. In the figure this residue is denoted by $\Delta l\equiv\Delta_{\rm PAM}=l_{s1}\pm l_{s2}-l_{s3}+m$ modulo 3. Only the $\Delta l=0$ class is compatible with a nonzero screw-symmetric cubic vertex. The distribution in Fig.~\ref{fig:2}(b) is nearly uniform among the three residue classes, so approximately two thirds of the kinematically allowed triplets fall into symmetry-forbidden residue classes. This is not an accidental exclusion of isolated processes; it is a symmetry partition of the three-phonon phase space.

The weighted phase space $W_3(\omega)$ in Fig.~\ref{fig:2}(c) demonstrates that the selection rule remains important after including occupations, broadened energy conservation, and the density of final states. We use 400~K for the microscopic plots because it is the upper end of the temperature range considered here. At this temperature anharmonic scattering is strongest and the relative conductivity enhancement is smallest, so the suppression is a conservative reference point. The reduction is strongest in the optical sector, where the PAM labels are most distinct, but it is also visible at lower frequencies. This already hints at the central transport mechanism: chiral optical modes do not need to carry the heat directly in order to control the scattering of the dominant low-frequency carriers.

A useful way to interpret Fig.~\ref{fig:2} is that PAM conservation changes the connectivity of the three-phonon scattering network before the mode lifetimes are evaluated. In the diagnostic PAM-relaxed baseline, a low-frequency heat-carrying mode can be coupled to many kinematically allowed triplets involving optical branches. In the PAM-conserving physical channel set, large groups of these acoustic--optical triplets are identified as symmetry-forbidden even though they satisfy the usual energy and crystal-momentum constraints. The remaining channels therefore form a different set of allowed final states. This distinction is important because thermal conductivity is controlled not only by the density of states near a given frequency, but also by how low-frequency heat carriers are connected to the rest of the spectrum through anharmonic scattering.

\begin{figure}[!tbp]
    \centering
    \includegraphics[width=0.9\columnwidth,height=0.48\textheight,keepaspectratio]{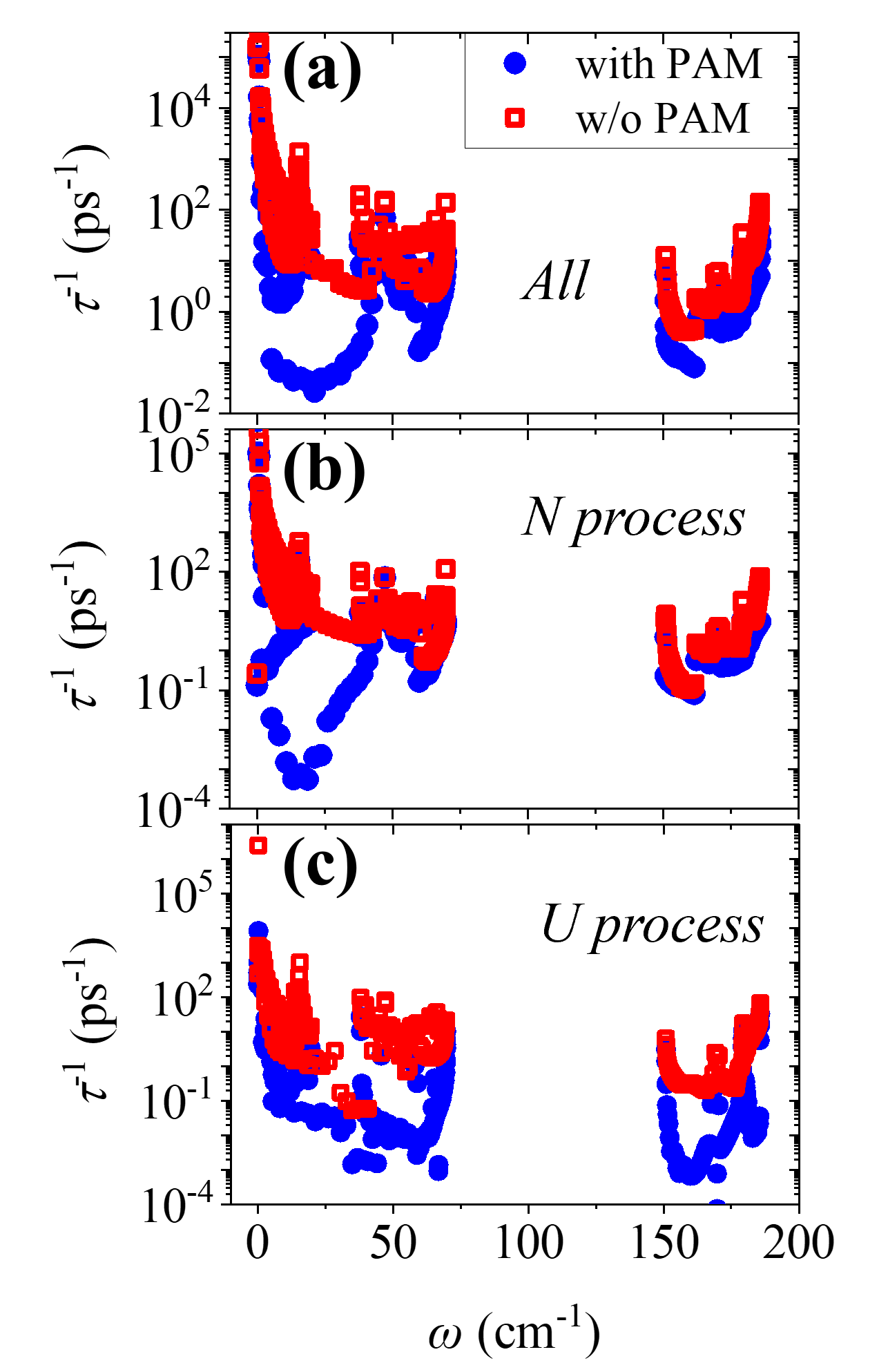}
    \caption{Suppression of microscopic scattering dynamics by PAM conservation. Frequency-resolved three-phonon scattering rates, shown as inverse lifetimes $\tau^{-1}$ at 400~K, for (a) all processes, (b) normal (N) processes, and (c) Umklapp (U) processes, comparing the PAM-conserving physical calculation with the diagnostic baseline labeled ``w/o PAM'' in the figure.}
    \label{fig:3}
\end{figure}

The phase-space reduction alone does not determine phonon lifetimes, because the three-phonon matrix elements and thermal occupations also enter the scattering rates. Figure~\ref{fig:3} therefore compares the frequency-resolved rates obtained from the PAM-conserving physical channel set with those from the diagnostic PAM-relaxed baseline. The PAM-conserving channel set yields lower total rates over broad frequency regions, confirming that the screw-PAM rule strongly reorganizes actual anharmonic relaxation rather than only the channel count. The reduction is not a uniform rescaling by one third: different branches have different third-order force constants, occupation factors, and access to final states, so the rate suppression is strongly mode and frequency dependent.

The normal/Umklapp decomposition makes the mechanism explicit: the PAM rule changes the allowed anharmonic network, and the resulting pruning preferentially removes resistive Umklapp pathways. Relative to the diagnostic baseline, normal processes are reduced over broad low- and intermediate-frequency regions, indicating that PAM conservation also reorganizes the redistribution of nonequilibrium phonons among modes. The Umklapp channel shows the stronger and more systematic suppression in Fig.~\ref{fig:3}(c). Many of the removed channels involve acoustic--optical scattering partners: visibly chiral optical modes therefore influence heat transport indirectly, by constraining relaxation pathways available to low-frequency heat carriers. Since Umklapp processes are the intrinsic momentum-relaxing events that directly relax the heat current, preferential suppression of resistive Umklapp pathways provides the microscopic origin of the conductivity enhancement discussed below.

\begin{figure}[!tbp]
    \centering
    \includegraphics[width=\columnwidth,height=0.50\textheight,keepaspectratio]{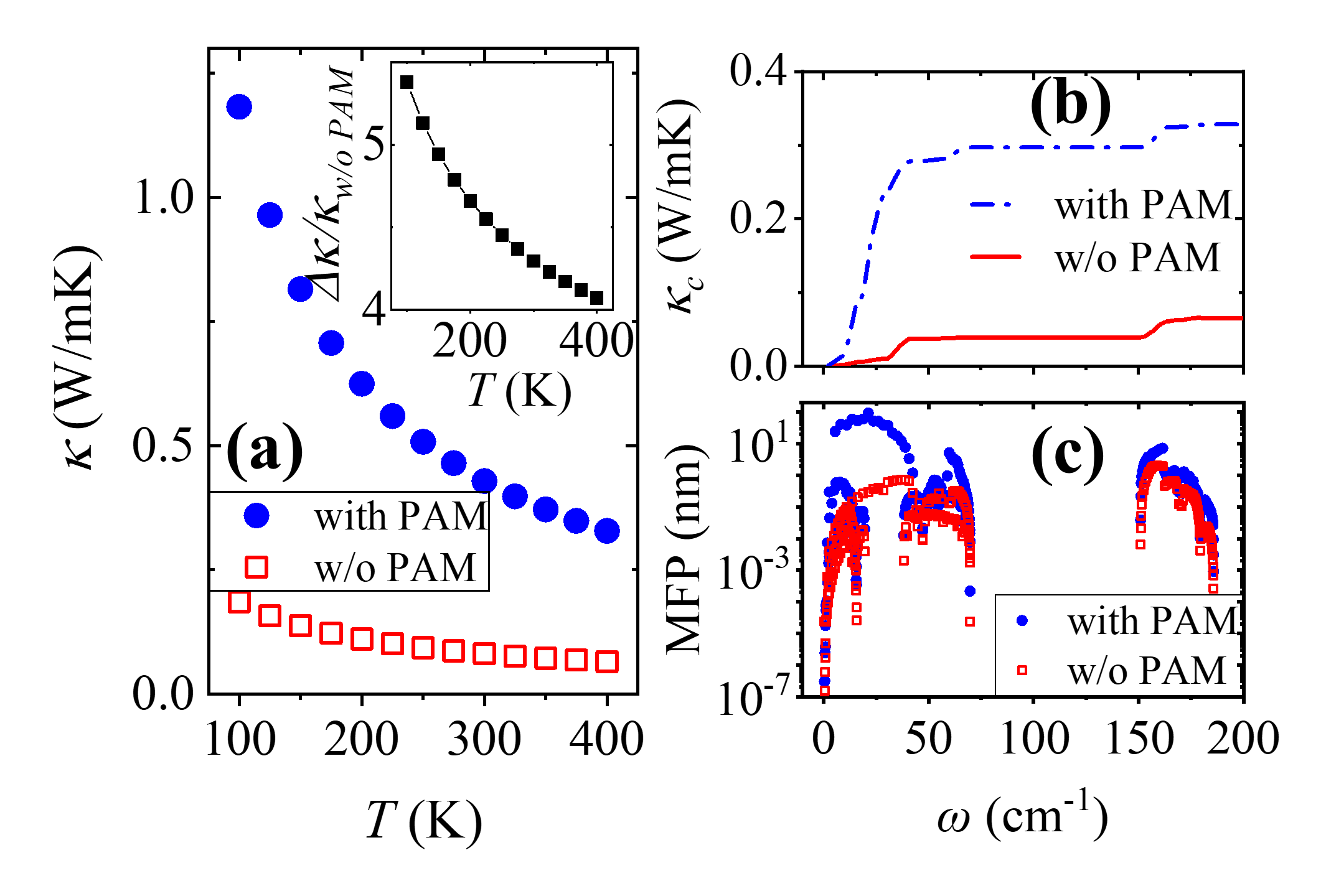}
    \caption{Thermal transport enhanced by PAM conservation. (a) Lattice thermal conductivity as a function of temperature; the inset shows the relative increase $\Delta\kappa/\kappa_{\rm w/o\;PAM}$, where ``w/o PAM'' denotes the diagnostic PAM-relaxed baseline defined in the text. (b) Cumulative thermal conductivity $\kappa_c$ as a function of frequency. (c) Mode-resolved phonon mean free path (MFP) as a function of frequency, with $\Lambda=|v_g^z|\tau^{\rm eff}$ and units of nm.}
    \label{fig:4}
\end{figure}

This scattering-network change produces the macroscopic consequence shown in Fig.~\ref{fig:4}(a). The PAM-conserving physical calculation enhances the lattice thermal conductivity throughout 100--400~K relative to the diagnostic PAM-relaxed baseline. The effect is order-one across the full temperature window: \(\kappa_{\rm PAM}/\kappa_{\rm w/o\;PAM}=6.38\), \(5.30\), and \(5.07\) at 100, 300, and 400~K, respectively. PAM conservation therefore remains an active transport constraint at the upper end of the temperature range considered here, where intrinsic phonon-phonon scattering is strongest.

Low-dimensional rotational systems already show order-one thermal-conductivity changes controlled by angular quantum numbers. Related line-group analyses of single-walled carbon nanotubes use angular quantum numbers to enumerate phonon-phonon channels and suppress resistive Umklapp relaxation~\cite{mahan2004flexure,mingo2005carbon,lindsay2009lattice,lindsay2010diameter}. In SWCNT bundles, rotational-symmetry breaking quenches symmetry-sensitive twist modes and strongly suppresses thermal conductivity~\cite{Shiga2024SWCNTBundling,Tao2026SWCNTBoseEinstein}. Related first-principles work has also connected phonon chirality to transport in one-dimensional and bulk Ba$_3$N-derived materials~\cite{pandey2018symmetry}. Our result is the complementary symmetry-preserving case: enforcing screw-PAM conservation removes resistive scattering pathways and enhances intrinsic thermal transport.

Figure~\ref{fig:4}(b,c) completes the mechanism chain by connecting the removed Umklapp pathways to low-frequency heat carriers. Although the optical modes display the clearest chiral character, the additional conductivity accumulates mainly in the low-frequency sector. This is not a contradiction. Optical chiral modes act as symmetry-sensitive scattering partners, while low-frequency modes provide the dominant heat current. Figure~\ref{fig:4}(c) gives the same conclusion in real-space language: in the PAM-conserving physical calculation, low-frequency phonons extend to longer mean free paths than in the diagnostic baseline, consistent with the suppressed Umklapp rates in Fig.~\ref{fig:3}. Thus the selection rule reshapes not only the number of scattering channels but also the frequency and length-scale structure of intrinsic heat transport.

We further tested this symmetry-resolved viewpoint in bulk $\alpha$-quartz, a chiral crystal for which first-principles Boltzmann calculations have been reported to underestimate measured lattice thermal conductivities despite neglecting defect scattering~\cite{Mizokami2018SiO2ThermalConductivity}. In a three-dimensional Brillouin zone, the exact scalar PAM rule is restricted to rotation/screw fixed lines. We therefore use $\alpha$-quartz as a finite-$q$ eigenphase-character benchmark tied to these high-symmetry manifolds. This PAM-explicit benchmark increases the calculated bulk thermal conductivity by about \(24\%\), comparable to the previously reported first-principles underestimation. We do not attribute the experimental discrepancy to PAM conservation alone; rather, the benchmark identifies a previously omitted symmetry-resolved organization of anharmonic scattering with the correct sign and comparable scale to alleviate this theory--experiment mismatch. Details are given in Sec.~S6 of the Supplemental Material~\cite{SM}.

This separation between ``where chirality is most visible'' and ``where heat is carried'' is central to the physical picture. A naive interpretation might expect optical chiral phonons to carry the thermal current directly. The calculations instead show a more collective mechanism: the symmetry classification of all branches constrains how optical and acoustic sectors exchange energy and momentum. The long-lived heat carriers are low-frequency modes, but their lifetimes are protected because screw-PAM-incompatible scattering partners are identified as symmetry-forbidden. The effect is therefore genuinely a property of the full anharmonic network rather than a branch-resolved contribution from optical modes alone.

The same symmetry mechanism also supports chirality-selected directional propagation. Thermal rectification was originally proposed in nonlinear lattices~\cite{li2004thermal} and demonstrated in nanoscale asymmetric systems~\cite{chang2006solid}; more recently, chiral-phonon diode concepts have been proposed in chiral crystals~\cite{chen2022chiral,li2024utilizing}. In the present chain, the key ingredient is PAM-velocity locking. Figure~\ref{fig:5}(a) shows optical windows in which the sign of $v_g^z$ is strongly correlated with $l_s$. A chirality-selective excitation can therefore preferentially populate right-moving or left-moving optical phonons. The ballistic fan in Fig.~\ref{fig:5}(b) illustrates the ideal handedness-selected propagation. In a finite device, scattering attenuates the selected population; Fig.~\ref{fig:5}(c) shows that the PAM-conserving lifetimes preserve a larger rectification estimate over the length range considered. The PAM-conserving physical calculation and diagnostic PAM-relaxed baseline labeled ``w/o PAM'' in Fig.~\ref{fig:5} use the same excitation windows and helicity-dependent weights, so the difference isolates the effect of screw-PAM-constrained phonon-phonon scattering.

\textit{Conclusion.--} We have shown that pseudo-angular momentum is not merely a spectroscopic selection-rule label for chiral phonons, but a symmetry quantum number that organizes intrinsic heat transport. In a $3_1$ helical Te chain, the PAM selection rule follows directly from screw-phase conservation in three-phonon scattering and gives the representation-level nonzero condition for the cubic anharmonic vertex. Making this rule explicit in first-principles Boltzmann transport removes about two thirds of the kinematically allowed triplets, preferentially suppresses resistive Umklapp scattering, and enhances the Te-chain lattice thermal conductivity by factors of \(6.38\), \(5.30\), and \(5.07\) at 100, 300, and 400~K. The enhancement is carried mainly by low-frequency, long-mean-free-path heat carriers, while visibly chiral optical modes act as symmetry-sensitive scattering partners that constrain their relaxation pathways. For the chiral optical branches, PAM-velocity locking further connects the enhanced lifetimes to chirality-selected directional propagation. Beyond this exact chain, the same eigenphase-conservation argument applies to high-symmetry phonon scattering in chiral materials with rotational or screw symmetry and to higher-order anharmonic interactions. The \(\alpha\)-quartz benchmark further yields an increase of about \(24\%\) in calculated bulk thermal conductivity, showing that symmetry-resolved scattering organization can still be physically significant at a material scale beyond the one-dimensional proof platform. These results establish PAM conservation as a transport selection principle for incoherent, finite-\(q\), thermally populated anharmonic scattering networks, showing that the same symmetry quantum number that enables optical addressability can also reorganize phonon relaxation, enhance thermal conductivity, and guide directional thermal transport.

\begin{acknowledgments}
This work was supported by the National Key Research and Development Program of China (No.~2023YFA1407001), Department of Science and Technology of Jiangsu Province (No.~BK20220032) and National Natural Science Foundation of China (No.~12504013).
\end{acknowledgments}

\nocite{feng2017four,Kim2026QuartzAcousticPAM,Komiyama2022ApproximateScrew,Li2024MoO3PAM,Nii2021MnSiFiniteQAngularMomentum,ueda2023chiral,yang2021crystal}
\bibliography{file}

\begin{thebibliography}{47}%
\makeatletter
\providecommand \@ifxundefined [1]{%
 \@ifx{#1\undefined}
}%
\providecommand \@ifnum [1]{%
 \ifnum #1\expandafter \@firstoftwo
 \else \expandafter \@secondoftwo
 \fi
}%
\providecommand \@ifx [1]{%
 \ifx #1\expandafter \@firstoftwo
 \else \expandafter \@secondoftwo
 \fi
}%
\providecommand \natexlab [1]{#1}%
\providecommand \enquote  [1]{``#1''}%
\providecommand \bibnamefont  [1]{#1}%
\providecommand \bibfnamefont [1]{#1}%
\providecommand \citenamefont [1]{#1}%
\providecommand \href@noop [0]{\@secondoftwo}%
\providecommand \href [0]{\begingroup \@sanitize@url \@href}%
\providecommand \@href[1]{\@@startlink{#1}\@@href}%
\providecommand \@@href[1]{\endgroup#1\@@endlink}%
\providecommand \@sanitize@url [0]{\catcode `\\12\catcode `\$12\catcode
  `\&12\catcode `\#12\catcode `\^12\catcode `\_12\catcode `\%12\relax}%
\providecommand \@@startlink[1]{}%
\providecommand \@@endlink[0]{}%
\providecommand \url  [0]{\begingroup\@sanitize@url \@url }%
\providecommand \@url [1]{\endgroup\@href {#1}{\urlprefix }}%
\providecommand \urlprefix  [0]{URL }%
\providecommand \Eprint [0]{\href }%
\providecommand \doibase [0]{https://doi.org/}%
\providecommand \selectlanguage [0]{\@gobble}%
\providecommand \bibinfo  [0]{\@secondoftwo}%
\providecommand \bibfield  [0]{\@secondoftwo}%
\providecommand \translation [1]{[#1]}%
\providecommand \BibitemOpen [0]{}%
\providecommand \bibitemStop [0]{}%
\providecommand \bibitemNoStop [0]{.\EOS\space}%
\providecommand \EOS [0]{\spacefactor3000\relax}%
\providecommand \BibitemShut  [1]{\csname bibitem#1\endcsname}%
\let\auto@bib@innerbib\@empty
\bibitem [{\citenamefont {Zhang}\ and\ \citenamefont
  {Niu}(2015)}]{zhang2015chiral}%
  \BibitemOpen
  \bibfield  {author} {\bibinfo {author} {\bibfnamefont {L.}~\bibnamefont
  {Zhang}}\ and\ \bibinfo {author} {\bibfnamefont {Q.}~\bibnamefont {Niu}},\
  }\href@noop {} {\bibfield  {journal} {\bibinfo  {journal} {Phys. Rev. Lett.}\
  }\textbf {\bibinfo {volume} {115}},\ \bibinfo {pages} {115502} (\bibinfo
  {year} {2015})}\BibitemShut {NoStop}%
\bibitem [{\citenamefont {Zhu}\ \emph {et~al.}(2018)\citenamefont {Zhu},
  \citenamefont {Yi}, \citenamefont {Li}, \citenamefont {Xiao}, \citenamefont
  {Zhang}, \citenamefont {Yang}, \citenamefont {Kaindl}, \citenamefont {Li},
  \citenamefont {Wang},\ and\ \citenamefont {Zhang}}]{zhu2018observation}%
  \BibitemOpen
  \bibfield  {author} {\bibinfo {author} {\bibfnamefont {H.}~\bibnamefont
  {Zhu}}, \bibinfo {author} {\bibfnamefont {J.}~\bibnamefont {Yi}}, \bibinfo
  {author} {\bibfnamefont {M.-Y.}\ \bibnamefont {Li}}, \bibinfo {author}
  {\bibfnamefont {J.}~\bibnamefont {Xiao}}, \bibinfo {author} {\bibfnamefont
  {L.}~\bibnamefont {Zhang}}, \bibinfo {author} {\bibfnamefont {C.-W.}\
  \bibnamefont {Yang}}, \bibinfo {author} {\bibfnamefont {R.~A.}\ \bibnamefont
  {Kaindl}}, \bibinfo {author} {\bibfnamefont {L.-J.}\ \bibnamefont {Li}},
  \bibinfo {author} {\bibfnamefont {Y.}~\bibnamefont {Wang}},\ and\ \bibinfo
  {author} {\bibfnamefont {X.}~\bibnamefont {Zhang}},\ }\href@noop {}
  {\bibfield  {journal} {\bibinfo  {journal} {Science}\ }\textbf {\bibinfo
  {volume} {359}},\ \bibinfo {pages} {579} (\bibinfo {year}
  {2018})}\BibitemShut {NoStop}%
\bibitem [{\citenamefont {Wang}\ \emph {et~al.}(2024)\citenamefont {Wang},
  \citenamefont {Sun}, \citenamefont {Li},\ and\ \citenamefont
  {Zhang}}]{wang2024chiral}%
  \BibitemOpen
  \bibfield  {author} {\bibinfo {author} {\bibfnamefont {T.}~\bibnamefont
  {Wang}}, \bibinfo {author} {\bibfnamefont {H.}~\bibnamefont {Sun}}, \bibinfo
  {author} {\bibfnamefont {X.}~\bibnamefont {Li}},\ and\ \bibinfo {author}
  {\bibfnamefont {L.}~\bibnamefont {Zhang}},\ }\href@noop {} {\bibfield
  {journal} {\bibinfo  {journal} {Nano Lett.}\ }\textbf {\bibinfo {volume}
  {24}},\ \bibinfo {pages} {4311} (\bibinfo {year} {2024})}\BibitemShut
  {NoStop}%
\bibitem [{\citenamefont {Juraschek}\ \emph {et~al.}(2025)\citenamefont
  {Juraschek}, \citenamefont {Geilhufe}, \citenamefont {Zhu}, \citenamefont
  {Basini}, \citenamefont {Baum}, \citenamefont {Baydin}, \citenamefont
  {Chaudhary}, \citenamefont {Fechner}, \citenamefont {Flebus}, \citenamefont
  {Grissonnanche}, \citenamefont {Kirilyuk}, \citenamefont {Lemeshko},
  \citenamefont {Maehrlein}, \citenamefont {Mignolet}, \citenamefont
  {Murakami}, \citenamefont {Niu}, \citenamefont {Nowak}, \citenamefont
  {Romao}, \citenamefont {Rostami}, \citenamefont {Satoh}, \citenamefont
  {Spaldin}, \citenamefont {Ueda},\ and\ \citenamefont
  {Zhang}}]{juraschek2025chiral}%
  \BibitemOpen
  \bibfield  {author} {\bibinfo {author} {\bibfnamefont {D.~M.}\ \bibnamefont
  {Juraschek}}, \bibinfo {author} {\bibfnamefont {R.~M.}\ \bibnamefont
  {Geilhufe}}, \bibinfo {author} {\bibfnamefont {H.}~\bibnamefont {Zhu}},
  \bibinfo {author} {\bibfnamefont {M.}~\bibnamefont {Basini}}, \bibinfo
  {author} {\bibfnamefont {P.}~\bibnamefont {Baum}}, \bibinfo {author}
  {\bibfnamefont {A.}~\bibnamefont {Baydin}}, \bibinfo {author} {\bibfnamefont
  {S.}~\bibnamefont {Chaudhary}}, \bibinfo {author} {\bibfnamefont
  {M.}~\bibnamefont {Fechner}}, \bibinfo {author} {\bibfnamefont
  {B.}~\bibnamefont {Flebus}}, \bibinfo {author} {\bibfnamefont
  {G.}~\bibnamefont {Grissonnanche}}, \bibinfo {author} {\bibfnamefont {A.~I.}\
  \bibnamefont {Kirilyuk}}, \bibinfo {author} {\bibfnamefont {M.}~\bibnamefont
  {Lemeshko}}, \bibinfo {author} {\bibfnamefont {S.~F.}\ \bibnamefont
  {Maehrlein}}, \bibinfo {author} {\bibfnamefont {M.}~\bibnamefont {Mignolet}},
  \bibinfo {author} {\bibfnamefont {S.}~\bibnamefont {Murakami}}, \bibinfo
  {author} {\bibfnamefont {Q.}~\bibnamefont {Niu}}, \bibinfo {author}
  {\bibfnamefont {U.}~\bibnamefont {Nowak}}, \bibinfo {author} {\bibfnamefont
  {C.~P.}\ \bibnamefont {Romao}}, \bibinfo {author} {\bibfnamefont
  {H.}~\bibnamefont {Rostami}}, \bibinfo {author} {\bibfnamefont
  {T.}~\bibnamefont {Satoh}}, \bibinfo {author} {\bibfnamefont {N.~A.}\
  \bibnamefont {Spaldin}}, \bibinfo {author} {\bibfnamefont {H.}~\bibnamefont
  {Ueda}},\ and\ \bibinfo {author} {\bibfnamefont {L.}~\bibnamefont {Zhang}},\
  }\href@noop {} {\bibfield  {journal} {\bibinfo  {journal} {Nat. Phys.}\
  }\textbf {\bibinfo {volume} {21}},\ \bibinfo {pages} {1532} (\bibinfo {year}
  {2025})}\BibitemShut {NoStop}%
\bibitem [{\citenamefont {Zhang}\ and\ \citenamefont
  {Niu}(2014)}]{zhang2014angular}%
  \BibitemOpen
  \bibfield  {author} {\bibinfo {author} {\bibfnamefont {L.}~\bibnamefont
  {Zhang}}\ and\ \bibinfo {author} {\bibfnamefont {Q.}~\bibnamefont {Niu}},\
  }\href@noop {} {\bibfield  {journal} {\bibinfo  {journal} {Phys. Rev. Lett.}\
  }\textbf {\bibinfo {volume} {112}},\ \bibinfo {pages} {085503} (\bibinfo
  {year} {2014})}\BibitemShut {NoStop}%
\bibitem [{\citenamefont {Juraschek}\ and\ \citenamefont
  {Spaldin}(2019)}]{juraschek2019orbital}%
  \BibitemOpen
  \bibfield  {author} {\bibinfo {author} {\bibfnamefont {D.~M.}\ \bibnamefont
  {Juraschek}}\ and\ \bibinfo {author} {\bibfnamefont {N.~A.}\ \bibnamefont
  {Spaldin}},\ }\href@noop {} {\bibfield  {journal} {\bibinfo  {journal} {Phys.
  Rev. Mater.}\ }\textbf {\bibinfo {volume} {3}},\ \bibinfo {pages} {064405}
  (\bibinfo {year} {2019})}\BibitemShut {NoStop}%
\bibitem [{\citenamefont {Kim}\ \emph {et~al.}(2023)\citenamefont {Kim},
  \citenamefont {Vetter}, \citenamefont {Yan}, \citenamefont {Yang},
  \citenamefont {Wang}, \citenamefont {Sun}, \citenamefont {Yang},
  \citenamefont {Comstock}, \citenamefont {Li}, \citenamefont {Zhou},
  \citenamefont {Zhang}, \citenamefont {You}, \citenamefont {Sun},\ and\
  \citenamefont {Liu}}]{kim2023chiral}%
  \BibitemOpen
  \bibfield  {author} {\bibinfo {author} {\bibfnamefont {K.}~\bibnamefont
  {Kim}}, \bibinfo {author} {\bibfnamefont {E.}~\bibnamefont {Vetter}},
  \bibinfo {author} {\bibfnamefont {L.}~\bibnamefont {Yan}}, \bibinfo {author}
  {\bibfnamefont {C.}~\bibnamefont {Yang}}, \bibinfo {author} {\bibfnamefont
  {Z.}~\bibnamefont {Wang}}, \bibinfo {author} {\bibfnamefont {R.}~\bibnamefont
  {Sun}}, \bibinfo {author} {\bibfnamefont {Y.}~\bibnamefont {Yang}}, \bibinfo
  {author} {\bibfnamefont {A.~H.}\ \bibnamefont {Comstock}}, \bibinfo {author}
  {\bibfnamefont {X.}~\bibnamefont {Li}}, \bibinfo {author} {\bibfnamefont
  {J.}~\bibnamefont {Zhou}}, \bibinfo {author} {\bibfnamefont {L.}~\bibnamefont
  {Zhang}}, \bibinfo {author} {\bibfnamefont {W.}~\bibnamefont {You}}, \bibinfo
  {author} {\bibfnamefont {D.}~\bibnamefont {Sun}},\ and\ \bibinfo {author}
  {\bibfnamefont {J.}~\bibnamefont {Liu}},\ }\href@noop {} {\bibfield
  {journal} {\bibinfo  {journal} {Nat. Mater.}\ }\textbf {\bibinfo {volume}
  {22}},\ \bibinfo {pages} {322} (\bibinfo {year} {2023})}\BibitemShut
  {NoStop}%
\bibitem [{\citenamefont {Wang}\ \emph {et~al.}(2022)\citenamefont {Wang},
  \citenamefont {Li}, \citenamefont {Zhu}, \citenamefont {Chen}, \citenamefont
  {Wu}, \citenamefont {Gao}, \citenamefont {Zhang},\ and\ \citenamefont
  {Yang}}]{wang2022chiral}%
  \BibitemOpen
  \bibfield  {author} {\bibinfo {author} {\bibfnamefont {Q.}~\bibnamefont
  {Wang}}, \bibinfo {author} {\bibfnamefont {S.}~\bibnamefont {Li}}, \bibinfo
  {author} {\bibfnamefont {J.}~\bibnamefont {Zhu}}, \bibinfo {author}
  {\bibfnamefont {H.}~\bibnamefont {Chen}}, \bibinfo {author} {\bibfnamefont
  {W.}~\bibnamefont {Wu}}, \bibinfo {author} {\bibfnamefont {W.}~\bibnamefont
  {Gao}}, \bibinfo {author} {\bibfnamefont {L.}~\bibnamefont {Zhang}},\ and\
  \bibinfo {author} {\bibfnamefont {S.~A.}\ \bibnamefont {Yang}},\ }\href@noop
  {} {\bibfield  {journal} {\bibinfo  {journal} {Phys. Rev. B}\ }\textbf
  {\bibinfo {volume} {105}},\ \bibinfo {pages} {104301} (\bibinfo {year}
  {2022})}\BibitemShut {NoStop}%
\bibitem [{\citenamefont {Pan}\ and\ \citenamefont
  {Caruso}(2023)}]{pan2023vibrational}%
  \BibitemOpen
  \bibfield  {author} {\bibinfo {author} {\bibfnamefont {Y.}~\bibnamefont
  {Pan}}\ and\ \bibinfo {author} {\bibfnamefont {F.}~\bibnamefont {Caruso}},\
  }\href@noop {} {\bibfield  {journal} {\bibinfo  {journal} {Nano Lett.}\
  }\textbf {\bibinfo {volume} {23}},\ \bibinfo {pages} {7463} (\bibinfo {year}
  {2023})}\BibitemShut {NoStop}%
\bibitem [{\citenamefont {Luo}\ \emph {et~al.}(2023)\citenamefont {Luo},
  \citenamefont {Lin}, \citenamefont {Zhang}, \citenamefont {Chen},
  \citenamefont {Blackert}, \citenamefont {Xu}, \citenamefont {Yakobson},\ and\
  \citenamefont {Zhu}}]{luo2023large}%
  \BibitemOpen
  \bibfield  {author} {\bibinfo {author} {\bibfnamefont {J.}~\bibnamefont
  {Luo}}, \bibinfo {author} {\bibfnamefont {T.}~\bibnamefont {Lin}}, \bibinfo
  {author} {\bibfnamefont {J.}~\bibnamefont {Zhang}}, \bibinfo {author}
  {\bibfnamefont {X.}~\bibnamefont {Chen}}, \bibinfo {author} {\bibfnamefont
  {E.~R.}\ \bibnamefont {Blackert}}, \bibinfo {author} {\bibfnamefont
  {R.}~\bibnamefont {Xu}}, \bibinfo {author} {\bibfnamefont {B.~I.}\
  \bibnamefont {Yakobson}},\ and\ \bibinfo {author} {\bibfnamefont
  {H.}~\bibnamefont {Zhu}},\ }\href@noop {} {\bibfield  {journal} {\bibinfo
  {journal} {Science}\ }\textbf {\bibinfo {volume} {382}},\ \bibinfo {pages}
  {698} (\bibinfo {year} {2023})}\BibitemShut {NoStop}%
\bibitem [{\citenamefont {Streib}(2021)}]{streib2021difference}%
  \BibitemOpen
  \bibfield  {author} {\bibinfo {author} {\bibfnamefont {S.}~\bibnamefont
  {Streib}},\ }\href@noop {} {\bibfield  {journal} {\bibinfo  {journal} {Phys.
  Rev. B}\ }\textbf {\bibinfo {volume} {103}},\ \bibinfo {pages} {L100409}
  (\bibinfo {year} {2021})}\BibitemShut {NoStop}%
\bibitem [{\citenamefont {Zhang}\ and\ \citenamefont
  {Murakami}(2022)}]{zhang2022chiral}%
  \BibitemOpen
  \bibfield  {author} {\bibinfo {author} {\bibfnamefont {T.}~\bibnamefont
  {Zhang}}\ and\ \bibinfo {author} {\bibfnamefont {S.}~\bibnamefont
  {Murakami}},\ }\href@noop {} {\bibfield  {journal} {\bibinfo  {journal}
  {Phys. Rev. Res.}\ }\textbf {\bibinfo {volume} {4}},\ \bibinfo {pages}
  {L012024} (\bibinfo {year} {2022})}\BibitemShut {NoStop}%
\bibitem [{\citenamefont {Ishito}\ \emph {et~al.}(2023)\citenamefont {Ishito},
  \citenamefont {Mao}, \citenamefont {Kobayashi}, \citenamefont {Kousaka},
  \citenamefont {Togawa}, \citenamefont {Kusunose}, \citenamefont {Kishine},\
  and\ \citenamefont {Satoh}}]{ishito2023chiral}%
  \BibitemOpen
  \bibfield  {author} {\bibinfo {author} {\bibfnamefont {K.}~\bibnamefont
  {Ishito}}, \bibinfo {author} {\bibfnamefont {H.}~\bibnamefont {Mao}},
  \bibinfo {author} {\bibfnamefont {K.}~\bibnamefont {Kobayashi}}, \bibinfo
  {author} {\bibfnamefont {Y.}~\bibnamefont {Kousaka}}, \bibinfo {author}
  {\bibfnamefont {Y.}~\bibnamefont {Togawa}}, \bibinfo {author} {\bibfnamefont
  {H.}~\bibnamefont {Kusunose}}, \bibinfo {author} {\bibfnamefont {J.-i.}\
  \bibnamefont {Kishine}},\ and\ \bibinfo {author} {\bibfnamefont
  {T.}~\bibnamefont {Satoh}},\ }\href@noop {} {\bibfield  {journal} {\bibinfo
  {journal} {Chirality}\ }\textbf {\bibinfo {volume} {35}},\ \bibinfo {pages}
  {338} (\bibinfo {year} {2023})}\BibitemShut {NoStop}%
\bibitem [{\citenamefont {Spirito}\ \emph {et~al.}(2024)\citenamefont
  {Spirito}, \citenamefont {Marras},\ and\ \citenamefont
  {Mart{\'\i}n-Garc{\'\i}a}}]{spirito2024lattice}%
  \BibitemOpen
  \bibfield  {author} {\bibinfo {author} {\bibfnamefont {D.}~\bibnamefont
  {Spirito}}, \bibinfo {author} {\bibfnamefont {S.}~\bibnamefont {Marras}},\
  and\ \bibinfo {author} {\bibfnamefont {B.}~\bibnamefont
  {Mart{\'\i}n-Garc{\'\i}a}},\ }\href@noop {} {\bibfield  {journal} {\bibinfo
  {journal} {J. Mater. Chem. C}\ }\textbf {\bibinfo {volume} {12}},\ \bibinfo
  {pages} {2544} (\bibinfo {year} {2024})}\BibitemShut {NoStop}%
\bibitem [{\citenamefont {Oishi}\ \emph {et~al.}(2024)\citenamefont {Oishi},
  \citenamefont {Fujii},\ and\ \citenamefont {Koreeda}}]{oishi2024selective}%
  \BibitemOpen
  \bibfield  {author} {\bibinfo {author} {\bibfnamefont {E.}~\bibnamefont
  {Oishi}}, \bibinfo {author} {\bibfnamefont {Y.}~\bibnamefont {Fujii}},\ and\
  \bibinfo {author} {\bibfnamefont {A.}~\bibnamefont {Koreeda}},\ }\href@noop
  {} {\bibfield  {journal} {\bibinfo  {journal} {Phys. Rev. B}\ }\textbf
  {\bibinfo {volume} {109}},\ \bibinfo {pages} {104306} (\bibinfo {year}
  {2024})}\BibitemShut {NoStop}%
\bibitem [{\citenamefont {Pine}\ and\ \citenamefont
  {Dresselhaus}(1971)}]{pine1971raman}%
  \BibitemOpen
  \bibfield  {author} {\bibinfo {author} {\bibfnamefont {A.}~\bibnamefont
  {Pine}}\ and\ \bibinfo {author} {\bibfnamefont {G.}~\bibnamefont
  {Dresselhaus}},\ }\href@noop {} {\bibfield  {journal} {\bibinfo  {journal}
  {Phys. Rev. B}\ }\textbf {\bibinfo {volume} {4}},\ \bibinfo {pages} {356}
  (\bibinfo {year} {1971})}\BibitemShut {NoStop}%
\bibitem [{\citenamefont {Medeiros}\ \emph {et~al.}(2017)\citenamefont
  {Medeiros}, \citenamefont {Marks}, \citenamefont {Wynn}, \citenamefont
  {Vasylenko}, \citenamefont {Ramasse}, \citenamefont {Quigley}, \citenamefont
  {Sloan},\ and\ \citenamefont {Morris}}]{medeiros2017single}%
  \BibitemOpen
  \bibfield  {author} {\bibinfo {author} {\bibfnamefont {P.~V.}\ \bibnamefont
  {Medeiros}}, \bibinfo {author} {\bibfnamefont {S.}~\bibnamefont {Marks}},
  \bibinfo {author} {\bibfnamefont {J.~M.}\ \bibnamefont {Wynn}}, \bibinfo
  {author} {\bibfnamefont {A.}~\bibnamefont {Vasylenko}}, \bibinfo {author}
  {\bibfnamefont {Q.~M.}\ \bibnamefont {Ramasse}}, \bibinfo {author}
  {\bibfnamefont {D.}~\bibnamefont {Quigley}}, \bibinfo {author} {\bibfnamefont
  {J.}~\bibnamefont {Sloan}},\ and\ \bibinfo {author} {\bibfnamefont {A.~J.}\
  \bibnamefont {Morris}},\ }\href@noop {} {\bibfield  {journal} {\bibinfo
  {journal} {ACS Nano}\ }\textbf {\bibinfo {volume} {11}},\ \bibinfo {pages}
  {6178} (\bibinfo {year} {2017})}\BibitemShut {NoStop}%
\bibitem [{\citenamefont {Poborchii}\ \emph {et~al.}(2023)\citenamefont
  {Poborchii}, \citenamefont {Fokin},\ and\ \citenamefont
  {Shklyaev}}]{poborchii2023optical}%
  \BibitemOpen
  \bibfield  {author} {\bibinfo {author} {\bibfnamefont {V.~V.}\ \bibnamefont
  {Poborchii}}, \bibinfo {author} {\bibfnamefont {A.~V.}\ \bibnamefont
  {Fokin}},\ and\ \bibinfo {author} {\bibfnamefont {A.~A.}\ \bibnamefont
  {Shklyaev}},\ }\href@noop {} {\bibfield  {journal} {\bibinfo  {journal}
  {Nanoscale Adv.}\ }\textbf {\bibinfo {volume} {5}},\ \bibinfo {pages} {220}
  (\bibinfo {year} {2023})}\BibitemShut {NoStop}%
\bibitem [{\citenamefont {Liu}\ \emph {et~al.}(2018)\citenamefont {Liu},
  \citenamefont {Hu}, \citenamefont {Caputo}, \citenamefont {Sun},
  \citenamefont {Li}, \citenamefont {Zhao},\ and\ \citenamefont
  {Ren}}]{liu2018allotropes}%
  \BibitemOpen
  \bibfield  {author} {\bibinfo {author} {\bibfnamefont {Y.}~\bibnamefont
  {Liu}}, \bibinfo {author} {\bibfnamefont {S.}~\bibnamefont {Hu}}, \bibinfo
  {author} {\bibfnamefont {R.}~\bibnamefont {Caputo}}, \bibinfo {author}
  {\bibfnamefont {K.}~\bibnamefont {Sun}}, \bibinfo {author} {\bibfnamefont
  {Y.}~\bibnamefont {Li}}, \bibinfo {author} {\bibfnamefont {G.}~\bibnamefont
  {Zhao}},\ and\ \bibinfo {author} {\bibfnamefont {W.}~\bibnamefont {Ren}},\
  }\href@noop {} {\bibfield  {journal} {\bibinfo  {journal} {RSC Adv.}\
  }\textbf {\bibinfo {volume} {8}},\ \bibinfo {pages} {39650} (\bibinfo {year}
  {2018})}\BibitemShut {NoStop}%
\bibitem [{\citenamefont {Calavalle}\ \emph {et~al.}(2022)\citenamefont
  {Calavalle}, \citenamefont {Su{\'a}rez-Rodr{\'i}guez}, \citenamefont
  {Mart{\'i}n-Garc{\'i}a}, \citenamefont {Johansson}, \citenamefont {Vaz},
  \citenamefont {Yang}, \citenamefont {Maznichenko}, \citenamefont {Ostanin},
  \citenamefont {Mateo-Alonso}, \citenamefont {Chuvilin}, \citenamefont
  {Mertig}, \citenamefont {Gobbi}, \citenamefont {Casanova},\ and\
  \citenamefont {Hueso}}]{Calavalle2022TeNanowire}%
  \BibitemOpen
  \bibfield  {author} {\bibinfo {author} {\bibfnamefont {F.}~\bibnamefont
  {Calavalle}}, \bibinfo {author} {\bibfnamefont {M.}~\bibnamefont
  {Su{\'a}rez-Rodr{\'i}guez}}, \bibinfo {author} {\bibfnamefont
  {B.}~\bibnamefont {Mart{\'i}n-Garc{\'i}a}}, \bibinfo {author} {\bibfnamefont
  {A.}~\bibnamefont {Johansson}}, \bibinfo {author} {\bibfnamefont {D.~C.}\
  \bibnamefont {Vaz}}, \bibinfo {author} {\bibfnamefont {H.}~\bibnamefont
  {Yang}}, \bibinfo {author} {\bibfnamefont {I.~V.}\ \bibnamefont
  {Maznichenko}}, \bibinfo {author} {\bibfnamefont {S.}~\bibnamefont
  {Ostanin}}, \bibinfo {author} {\bibfnamefont {A.}~\bibnamefont
  {Mateo-Alonso}}, \bibinfo {author} {\bibfnamefont {A.}~\bibnamefont
  {Chuvilin}}, \bibinfo {author} {\bibfnamefont {I.}~\bibnamefont {Mertig}},
  \bibinfo {author} {\bibfnamefont {M.}~\bibnamefont {Gobbi}}, \bibinfo
  {author} {\bibfnamefont {F.}~\bibnamefont {Casanova}},\ and\ \bibinfo
  {author} {\bibfnamefont {L.~E.}\ \bibnamefont {Hueso}},\ }\href
  {https://doi.org/10.1038/s41563-022-01211-7} {\bibfield  {journal} {\bibinfo
  {journal} {Nat. Mater.}\ }\textbf {\bibinfo {volume} {21}},\ \bibinfo {pages}
  {526} (\bibinfo {year} {2022})}\BibitemShut {NoStop}%
\bibitem [{\citenamefont {Mizokami}\ \emph {et~al.}(2018)\citenamefont
  {Mizokami}, \citenamefont {Togo},\ and\ \citenamefont
  {Tanaka}}]{Mizokami2018SiO2ThermalConductivity}%
  \BibitemOpen
  \bibfield  {author} {\bibinfo {author} {\bibfnamefont {K.}~\bibnamefont
  {Mizokami}}, \bibinfo {author} {\bibfnamefont {A.}~\bibnamefont {Togo}},\
  and\ \bibinfo {author} {\bibfnamefont {I.}~\bibnamefont {Tanaka}},\
  }\href@noop {} {\bibfield  {journal} {\bibinfo  {journal} {Phys. Rev. B}\
  }\textbf {\bibinfo {volume} {97}},\ \bibinfo {pages} {224306} (\bibinfo
  {year} {2018})}\BibitemShut {NoStop}%
\bibitem [{\citenamefont {Peng}\ \emph {et~al.}(2015)\citenamefont {Peng},
  \citenamefont {Kioussis},\ and\ \citenamefont
  {Stewart}}]{peng2015anisotropic}%
  \BibitemOpen
  \bibfield  {author} {\bibinfo {author} {\bibfnamefont {H.}~\bibnamefont
  {Peng}}, \bibinfo {author} {\bibfnamefont {N.}~\bibnamefont {Kioussis}},\
  and\ \bibinfo {author} {\bibfnamefont {D.~A.}\ \bibnamefont {Stewart}},\
  }\href@noop {} {\bibfield  {journal} {\bibinfo  {journal} {Appl. Phys.
  Lett.}\ }\textbf {\bibinfo {volume} {107}},\ \bibinfo {pages} {251904}
  (\bibinfo {year} {2015})}\BibitemShut {NoStop}%
\bibitem [{\citenamefont {Ho}\ \emph {et~al.}(1972)\citenamefont {Ho},
  \citenamefont {Powell},\ and\ \citenamefont
  {Liley}}]{Ho1972ThermalConductivityElements}%
  \BibitemOpen
  \bibfield  {author} {\bibinfo {author} {\bibfnamefont {C.~Y.}\ \bibnamefont
  {Ho}}, \bibinfo {author} {\bibfnamefont {R.~W.}\ \bibnamefont {Powell}},\
  and\ \bibinfo {author} {\bibfnamefont {P.~E.}\ \bibnamefont {Liley}},\
  }\href@noop {} {\bibfield  {journal} {\bibinfo  {journal} {J. Phys. Chem.
  Ref. Data}\ }\textbf {\bibinfo {volume} {1}},\ \bibinfo {pages} {279}
  (\bibinfo {year} {1972})}\BibitemShut {NoStop}%
\bibitem [{SM()}]{SM}%
  \BibitemOpen
  \href@noop {} {}\bibinfo {note} {See Supplemental Material at [URL will be
  inserted by publisher] for derivations of the screw-PAM selection rule,
  computational details, definitions of the transport diagnostics, the
  thermal-rectification estimate, and the bulk alpha-quartz benchmark, which
  includes
  Refs.~\cite{feng2017four,Kim2026QuartzAcousticPAM,Komiyama2022ApproximateScrew,Li2024MoO3PAM,Nii2021MnSiFiniteQAngularMomentum,ueda2023chiral,yang2021crystal}.}\BibitemShut
  {Stop}%
\bibitem [{\citenamefont {Kresse}\ and\ \citenamefont
  {Furthm{\"u}ller}(1996)}]{kresse1996efficient}%
  \BibitemOpen
  \bibfield  {author} {\bibinfo {author} {\bibfnamefont {G.}~\bibnamefont
  {Kresse}}\ and\ \bibinfo {author} {\bibfnamefont {J.}~\bibnamefont
  {Furthm{\"u}ller}},\ }\href@noop {} {\bibfield  {journal} {\bibinfo
  {journal} {Phys. Rev. B}\ }\textbf {\bibinfo {volume} {54}},\ \bibinfo
  {pages} {11169} (\bibinfo {year} {1996})}\BibitemShut {NoStop}%
\bibitem [{\citenamefont {Gonze}\ and\ \citenamefont
  {Lee}(1997)}]{gonze1997dynamical}%
  \BibitemOpen
  \bibfield  {author} {\bibinfo {author} {\bibfnamefont {X.}~\bibnamefont
  {Gonze}}\ and\ \bibinfo {author} {\bibfnamefont {C.}~\bibnamefont {Lee}},\
  }\href@noop {} {\bibfield  {journal} {\bibinfo  {journal} {Phys. Rev. B}\
  }\textbf {\bibinfo {volume} {55}},\ \bibinfo {pages} {10355} (\bibinfo {year}
  {1997})}\BibitemShut {NoStop}%
\bibitem [{\citenamefont {Togo}\ \emph {et~al.}(2023)\citenamefont {Togo},
  \citenamefont {Chaput}, \citenamefont {Tadano},\ and\ \citenamefont
  {Tanaka}}]{phonopy-phono3py-JPCM}%
  \BibitemOpen
  \bibfield  {author} {\bibinfo {author} {\bibfnamefont {A.}~\bibnamefont
  {Togo}}, \bibinfo {author} {\bibfnamefont {L.}~\bibnamefont {Chaput}},
  \bibinfo {author} {\bibfnamefont {T.}~\bibnamefont {Tadano}},\ and\ \bibinfo
  {author} {\bibfnamefont {I.}~\bibnamefont {Tanaka}},\ }\href@noop {}
  {\bibfield  {journal} {\bibinfo  {journal} {J. Phys. Condens. Matter}\
  }\textbf {\bibinfo {volume} {35}},\ \bibinfo {pages} {353001} (\bibinfo
  {year} {2023})}\BibitemShut {NoStop}%
\bibitem [{\citenamefont {Eriksson}\ \emph {et~al.}(2019)\citenamefont
  {Eriksson}, \citenamefont {Fransson},\ and\ \citenamefont
  {Erhart}}]{eriksson2019hiphive}%
  \BibitemOpen
  \bibfield  {author} {\bibinfo {author} {\bibfnamefont {F.}~\bibnamefont
  {Eriksson}}, \bibinfo {author} {\bibfnamefont {E.}~\bibnamefont {Fransson}},\
  and\ \bibinfo {author} {\bibfnamefont {P.}~\bibnamefont {Erhart}},\
  }\href@noop {} {\bibfield  {journal} {\bibinfo  {journal} {Adv. Theory
  Simul.}\ }\textbf {\bibinfo {volume} {2}},\ \bibinfo {pages} {1800184}
  (\bibinfo {year} {2019})}\BibitemShut {NoStop}%
\bibitem [{\citenamefont {Li}\ \emph {et~al.}(2014)\citenamefont {Li},
  \citenamefont {Carrete}, \citenamefont {Katcho},\ and\ \citenamefont
  {Mingo}}]{li2014shengbte}%
  \BibitemOpen
  \bibfield  {author} {\bibinfo {author} {\bibfnamefont {W.}~\bibnamefont
  {Li}}, \bibinfo {author} {\bibfnamefont {J.}~\bibnamefont {Carrete}},
  \bibinfo {author} {\bibfnamefont {N.~A.}\ \bibnamefont {Katcho}},\ and\
  \bibinfo {author} {\bibfnamefont {N.}~\bibnamefont {Mingo}},\ }\href@noop {}
  {\bibfield  {journal} {\bibinfo  {journal} {Comput. Phys. Commun.}\ }\textbf
  {\bibinfo {volume} {185}},\ \bibinfo {pages} {1747} (\bibinfo {year}
  {2014})}\BibitemShut {NoStop}%
\bibitem [{\citenamefont {Mahan}\ and\ \citenamefont
  {Jeon}(2004)}]{mahan2004flexure}%
  \BibitemOpen
  \bibfield  {author} {\bibinfo {author} {\bibfnamefont {G.}~\bibnamefont
  {Mahan}}\ and\ \bibinfo {author} {\bibfnamefont {G.~S.}\ \bibnamefont
  {Jeon}},\ }\href@noop {} {\bibfield  {journal} {\bibinfo  {journal} {Phys.
  Rev. B}\ }\textbf {\bibinfo {volume} {70}},\ \bibinfo {pages} {075405}
  (\bibinfo {year} {2004})}\BibitemShut {NoStop}%
\bibitem [{\citenamefont {Mingo}\ and\ \citenamefont
  {Broido}(2005)}]{mingo2005carbon}%
  \BibitemOpen
  \bibfield  {author} {\bibinfo {author} {\bibfnamefont {N.}~\bibnamefont
  {Mingo}}\ and\ \bibinfo {author} {\bibfnamefont {D.}~\bibnamefont {Broido}},\
  }\href@noop {} {\bibfield  {journal} {\bibinfo  {journal} {Phys. Rev. Lett.}\
  }\textbf {\bibinfo {volume} {95}},\ \bibinfo {pages} {096105} (\bibinfo
  {year} {2005})}\BibitemShut {NoStop}%
\bibitem [{\citenamefont {Lindsay}\ \emph {et~al.}(2009)\citenamefont
  {Lindsay}, \citenamefont {Broido},\ and\ \citenamefont
  {Mingo}}]{lindsay2009lattice}%
  \BibitemOpen
  \bibfield  {author} {\bibinfo {author} {\bibfnamefont {L.}~\bibnamefont
  {Lindsay}}, \bibinfo {author} {\bibfnamefont {D.}~\bibnamefont {Broido}},\
  and\ \bibinfo {author} {\bibfnamefont {N.}~\bibnamefont {Mingo}},\
  }\href@noop {} {\bibfield  {journal} {\bibinfo  {journal} {Phys. Rev. B}\
  }\textbf {\bibinfo {volume} {80}},\ \bibinfo {pages} {125407} (\bibinfo
  {year} {2009})}\BibitemShut {NoStop}%
\bibitem [{\citenamefont {Lindsay}\ \emph {et~al.}(2010)\citenamefont
  {Lindsay}, \citenamefont {Broido},\ and\ \citenamefont
  {Mingo}}]{lindsay2010diameter}%
  \BibitemOpen
  \bibfield  {author} {\bibinfo {author} {\bibfnamefont {L.}~\bibnamefont
  {Lindsay}}, \bibinfo {author} {\bibfnamefont {D.}~\bibnamefont {Broido}},\
  and\ \bibinfo {author} {\bibfnamefont {N.}~\bibnamefont {Mingo}},\
  }\href@noop {} {\bibfield  {journal} {\bibinfo  {journal} {Phys. Rev. B}\
  }\textbf {\bibinfo {volume} {82}},\ \bibinfo {pages} {161402} (\bibinfo
  {year} {2010})}\BibitemShut {NoStop}%
\bibitem [{\citenamefont {Shiga}\ \emph {et~al.}(2024)\citenamefont {Shiga},
  \citenamefont {Terada}, \citenamefont {Chiashi},\ and\ \citenamefont
  {Kodama}}]{Shiga2024SWCNTBundling}%
  \BibitemOpen
  \bibfield  {author} {\bibinfo {author} {\bibfnamefont {T.}~\bibnamefont
  {Shiga}}, \bibinfo {author} {\bibfnamefont {Y.}~\bibnamefont {Terada}},
  \bibinfo {author} {\bibfnamefont {S.}~\bibnamefont {Chiashi}},\ and\ \bibinfo
  {author} {\bibfnamefont {T.}~\bibnamefont {Kodama}},\ }\href@noop {}
  {\bibfield  {journal} {\bibinfo  {journal} {Carbon}\ }\textbf {\bibinfo
  {volume} {223}},\ \bibinfo {pages} {119048} (\bibinfo {year}
  {2024})}\BibitemShut {NoStop}%
\bibitem [{\citenamefont {Tao}\ \emph {et~al.}(2026)\citenamefont {Tao},
  \citenamefont {Zhang}, \citenamefont {Tang}, \citenamefont {Maruyama},\ and\
  \citenamefont {Feng}}]{Tao2026SWCNTBoseEinstein}%
  \BibitemOpen
  \bibfield  {author} {\bibinfo {author} {\bibfnamefont {F.}~\bibnamefont
  {Tao}}, \bibinfo {author} {\bibfnamefont {X.}~\bibnamefont {Zhang}}, \bibinfo
  {author} {\bibfnamefont {D.}~\bibnamefont {Tang}}, \bibinfo {author}
  {\bibfnamefont {S.}~\bibnamefont {Maruyama}},\ and\ \bibinfo {author}
  {\bibfnamefont {Y.}~\bibnamefont {Feng}},\ }\href@noop {} {\bibfield
  {journal} {\bibinfo  {journal} {Nano Lett.}\ }\textbf {\bibinfo {volume}
  {26}},\ \bibinfo {pages} {3814} (\bibinfo {year} {2026})}\BibitemShut
  {NoStop}%
\bibitem [{\citenamefont {Pandey}\ \emph {et~al.}(2018)\citenamefont {Pandey},
  \citenamefont {Polanco}, \citenamefont {Cooper}, \citenamefont {Parker},\
  and\ \citenamefont {Lindsay}}]{pandey2018symmetry}%
  \BibitemOpen
  \bibfield  {author} {\bibinfo {author} {\bibfnamefont {T.}~\bibnamefont
  {Pandey}}, \bibinfo {author} {\bibfnamefont {C.~A.}\ \bibnamefont {Polanco}},
  \bibinfo {author} {\bibfnamefont {V.~R.}\ \bibnamefont {Cooper}}, \bibinfo
  {author} {\bibfnamefont {D.~S.}\ \bibnamefont {Parker}},\ and\ \bibinfo
  {author} {\bibfnamefont {L.}~\bibnamefont {Lindsay}},\ }\href@noop {}
  {\bibfield  {journal} {\bibinfo  {journal} {Phys. Rev. B}\ }\textbf {\bibinfo
  {volume} {98}},\ \bibinfo {pages} {241405} (\bibinfo {year}
  {2018})}\BibitemShut {NoStop}%
\bibitem [{\citenamefont {Li}\ \emph {et~al.}(2004)\citenamefont {Li},
  \citenamefont {Wang},\ and\ \citenamefont {Casati}}]{li2004thermal}%
  \BibitemOpen
  \bibfield  {author} {\bibinfo {author} {\bibfnamefont {B.}~\bibnamefont
  {Li}}, \bibinfo {author} {\bibfnamefont {L.}~\bibnamefont {Wang}},\ and\
  \bibinfo {author} {\bibfnamefont {G.}~\bibnamefont {Casati}},\ }\href@noop {}
  {\bibfield  {journal} {\bibinfo  {journal} {Phys. Rev. Lett.}\ }\textbf
  {\bibinfo {volume} {93}},\ \bibinfo {pages} {184301} (\bibinfo {year}
  {2004})}\BibitemShut {NoStop}%
\bibitem [{\citenamefont {Chang}\ \emph {et~al.}(2006)\citenamefont {Chang},
  \citenamefont {Okawa}, \citenamefont {Majumdar},\ and\ \citenamefont
  {Zettl}}]{chang2006solid}%
  \BibitemOpen
  \bibfield  {author} {\bibinfo {author} {\bibfnamefont {C.~W.}\ \bibnamefont
  {Chang}}, \bibinfo {author} {\bibfnamefont {D.}~\bibnamefont {Okawa}},
  \bibinfo {author} {\bibfnamefont {A.}~\bibnamefont {Majumdar}},\ and\
  \bibinfo {author} {\bibfnamefont {A.}~\bibnamefont {Zettl}},\ }\href@noop {}
  {\bibfield  {journal} {\bibinfo  {journal} {Science}\ }\textbf {\bibinfo
  {volume} {314}},\ \bibinfo {pages} {1121} (\bibinfo {year}
  {2006})}\BibitemShut {NoStop}%
\bibitem [{\citenamefont {Chen}\ \emph {et~al.}(2022)\citenamefont {Chen},
  \citenamefont {Wu}, \citenamefont {Zhu}, \citenamefont {Yang}, \citenamefont
  {Gong}, \citenamefont {Gao}, \citenamefont {Yang},\ and\ \citenamefont
  {Zhang}}]{chen2022chiral}%
  \BibitemOpen
  \bibfield  {author} {\bibinfo {author} {\bibfnamefont {H.}~\bibnamefont
  {Chen}}, \bibinfo {author} {\bibfnamefont {W.}~\bibnamefont {Wu}}, \bibinfo
  {author} {\bibfnamefont {J.}~\bibnamefont {Zhu}}, \bibinfo {author}
  {\bibfnamefont {Z.}~\bibnamefont {Yang}}, \bibinfo {author} {\bibfnamefont
  {W.}~\bibnamefont {Gong}}, \bibinfo {author} {\bibfnamefont {W.}~\bibnamefont
  {Gao}}, \bibinfo {author} {\bibfnamefont {S.~A.}\ \bibnamefont {Yang}},\ and\
  \bibinfo {author} {\bibfnamefont {L.}~\bibnamefont {Zhang}},\ }\href@noop {}
  {\bibfield  {journal} {\bibinfo  {journal} {Nano Lett.}\ }\textbf {\bibinfo
  {volume} {22}},\ \bibinfo {pages} {1688} (\bibinfo {year}
  {2022})}\BibitemShut {NoStop}%
\bibitem [{\citenamefont {Li}\ \emph {et~al.}(2024{\natexlab{a}})\citenamefont
  {Li}, \citenamefont {Long}, \citenamefont {Wang}, \citenamefont {Zhou},\ and\
  \citenamefont {Zhang}}]{li2024utilizing}%
  \BibitemOpen
  \bibfield  {author} {\bibinfo {author} {\bibfnamefont {X.}~\bibnamefont
  {Li}}, \bibinfo {author} {\bibfnamefont {Y.}~\bibnamefont {Long}}, \bibinfo
  {author} {\bibfnamefont {T.}~\bibnamefont {Wang}}, \bibinfo {author}
  {\bibfnamefont {Y.}~\bibnamefont {Zhou}},\ and\ \bibinfo {author}
  {\bibfnamefont {L.}~\bibnamefont {Zhang}},\ }\href@noop {} {\bibfield
  {journal} {\bibinfo  {journal} {Appl. Phys. Lett.}\ }\textbf {\bibinfo
  {volume} {124}},\ \bibinfo {pages} {252201} (\bibinfo {year}
  {2024}{\natexlab{a}})}\BibitemShut {NoStop}%
\bibitem [{\citenamefont {Feng}\ \emph {et~al.}(2017)\citenamefont {Feng},
  \citenamefont {Lindsay},\ and\ \citenamefont {Ruan}}]{feng2017four}%
  \BibitemOpen
  \bibfield  {author} {\bibinfo {author} {\bibfnamefont {T.}~\bibnamefont
  {Feng}}, \bibinfo {author} {\bibfnamefont {L.}~\bibnamefont {Lindsay}},\ and\
  \bibinfo {author} {\bibfnamefont {X.}~\bibnamefont {Ruan}},\ }\href@noop {}
  {\bibfield  {journal} {\bibinfo  {journal} {Phys. Rev. B}\ }\textbf {\bibinfo
  {volume} {96}},\ \bibinfo {pages} {161201} (\bibinfo {year}
  {2017})}\BibitemShut {NoStop}%
\bibitem [{\citenamefont {Kim}\ \emph {et~al.}(2026)\citenamefont {Kim},
  \citenamefont {Hwang}, \citenamefont {Moon}, \citenamefont {An},
  \citenamefont {Lee}, \citenamefont {Ko}, \citenamefont {Park}, \citenamefont
  {Choi},\ and\ \citenamefont {Hwang}}]{Kim2026QuartzAcousticPAM}%
  \BibitemOpen
  \bibfield  {author} {\bibinfo {author} {\bibfnamefont {C.}~\bibnamefont
  {Kim}}, \bibinfo {author} {\bibfnamefont {I.-K.}\ \bibnamefont {Hwang}},
  \bibinfo {author} {\bibfnamefont {K.-W.}\ \bibnamefont {Moon}}, \bibinfo
  {author} {\bibfnamefont {K.}~\bibnamefont {An}}, \bibinfo {author}
  {\bibfnamefont {K.-J.}\ \bibnamefont {Lee}}, \bibinfo {author} {\bibfnamefont
  {J.-H.}\ \bibnamefont {Ko}}, \bibinfo {author} {\bibfnamefont {B.-G.}\
  \bibnamefont {Park}}, \bibinfo {author} {\bibfnamefont {K.-Y.}\ \bibnamefont
  {Choi}},\ and\ \bibinfo {author} {\bibfnamefont {C.}~\bibnamefont {Hwang}},\
  }\href@noop {} {\bibfield  {journal} {\bibinfo  {journal} {Adv. Mater.}\
  }\textbf {\bibinfo {volume} {38}},\ \bibinfo {pages} {e11289} (\bibinfo
  {year} {2026})}\BibitemShut {NoStop}%
\bibitem [{\citenamefont {Komiyama}\ \emph {et~al.}(2022)\citenamefont
  {Komiyama}, \citenamefont {Zhang},\ and\ \citenamefont
  {Murakami}}]{Komiyama2022ApproximateScrew}%
  \BibitemOpen
  \bibfield  {author} {\bibinfo {author} {\bibfnamefont {H.}~\bibnamefont
  {Komiyama}}, \bibinfo {author} {\bibfnamefont {T.}~\bibnamefont {Zhang}},\
  and\ \bibinfo {author} {\bibfnamefont {S.}~\bibnamefont {Murakami}},\
  }\href@noop {} {\bibfield  {journal} {\bibinfo  {journal} {Phys. Rev. B}\
  }\textbf {\bibinfo {volume} {106}},\ \bibinfo {pages} {184104} (\bibinfo
  {year} {2022})}\BibitemShut {NoStop}%
\bibitem [{\citenamefont {Li}\ \emph {et~al.}(2024{\natexlab{b}})\citenamefont
  {Li}, \citenamefont {Li}, \citenamefont {Chen},\ and\ \citenamefont
  {Wang}}]{Li2024MoO3PAM}%
  \BibitemOpen
  \bibfield  {author} {\bibinfo {author} {\bibfnamefont {M.}~\bibnamefont
  {Li}}, \bibinfo {author} {\bibfnamefont {Z.}~\bibnamefont {Li}}, \bibinfo
  {author} {\bibfnamefont {H.}~\bibnamefont {Chen}},\ and\ \bibinfo {author}
  {\bibfnamefont {W.}~\bibnamefont {Wang}},\ }\href@noop {} {\bibfield
  {journal} {\bibinfo  {journal} {Nanomaterials}\ }\textbf {\bibinfo {volume}
  {14}},\ \bibinfo {pages} {607} (\bibinfo {year}
  {2024}{\natexlab{b}})}\BibitemShut {NoStop}%
\bibitem [{\citenamefont {Nii}\ \emph {et~al.}(2021)\citenamefont {Nii},
  \citenamefont {Hirokane}, \citenamefont {Koretsune}, \citenamefont
  {Ishikawa}, \citenamefont {Baron},\ and\ \citenamefont
  {Onose}}]{Nii2021MnSiFiniteQAngularMomentum}%
  \BibitemOpen
  \bibfield  {author} {\bibinfo {author} {\bibfnamefont {Y.}~\bibnamefont
  {Nii}}, \bibinfo {author} {\bibfnamefont {Y.}~\bibnamefont {Hirokane}},
  \bibinfo {author} {\bibfnamefont {T.}~\bibnamefont {Koretsune}}, \bibinfo
  {author} {\bibfnamefont {D.}~\bibnamefont {Ishikawa}}, \bibinfo {author}
  {\bibfnamefont {A.~Q.~R.}\ \bibnamefont {Baron}},\ and\ \bibinfo {author}
  {\bibfnamefont {Y.}~\bibnamefont {Onose}},\ }\href@noop {} {\bibfield
  {journal} {\bibinfo  {journal} {Phys. Rev. B}\ }\textbf {\bibinfo {volume}
  {104}},\ \bibinfo {pages} {L081101} (\bibinfo {year} {2021})}\BibitemShut
  {NoStop}%
\bibitem [{\citenamefont {Ueda}\ \emph {et~al.}(2023)\citenamefont {Ueda},
  \citenamefont {Garc{\'\i}a-Fern{\'a}ndez}, \citenamefont {Agrestini},
  \citenamefont {Romao}, \citenamefont {van~den Brink}, \citenamefont
  {Spaldin}, \citenamefont {Zhou},\ and\ \citenamefont
  {Staub}}]{ueda2023chiral}%
  \BibitemOpen
  \bibfield  {author} {\bibinfo {author} {\bibfnamefont {H.}~\bibnamefont
  {Ueda}}, \bibinfo {author} {\bibfnamefont {M.}~\bibnamefont
  {Garc{\'\i}a-Fern{\'a}ndez}}, \bibinfo {author} {\bibfnamefont
  {S.}~\bibnamefont {Agrestini}}, \bibinfo {author} {\bibfnamefont {C.~P.}\
  \bibnamefont {Romao}}, \bibinfo {author} {\bibfnamefont {J.}~\bibnamefont
  {van~den Brink}}, \bibinfo {author} {\bibfnamefont {N.~A.}\ \bibnamefont
  {Spaldin}}, \bibinfo {author} {\bibfnamefont {K.-J.}\ \bibnamefont {Zhou}},\
  and\ \bibinfo {author} {\bibfnamefont {U.}~\bibnamefont {Staub}},\
  }\href@noop {} {\bibfield  {journal} {\bibinfo  {journal} {Nature}\ }\textbf
  {\bibinfo {volume} {618}},\ \bibinfo {pages} {946} (\bibinfo {year}
  {2023})}\BibitemShut {NoStop}%
\bibitem [{\citenamefont {Yang}\ \emph {et~al.}(2021)\citenamefont {Yang},
  \citenamefont {Yue}, \citenamefont {Quan},\ and\ \citenamefont
  {Liao}}]{yang2021crystal}%
  \BibitemOpen
  \bibfield  {author} {\bibinfo {author} {\bibfnamefont {R.}~\bibnamefont
  {Yang}}, \bibinfo {author} {\bibfnamefont {S.}~\bibnamefont {Yue}}, \bibinfo
  {author} {\bibfnamefont {Y.}~\bibnamefont {Quan}},\ and\ \bibinfo {author}
  {\bibfnamefont {B.}~\bibnamefont {Liao}},\ }\href@noop {} {\bibfield
  {journal} {\bibinfo  {journal} {Phys. Rev. B}\ }\textbf {\bibinfo {volume}
  {103}},\ \bibinfo {pages} {184302} (\bibinfo {year} {2021})}\BibitemShut
  {NoStop}%
\end{thebibliography}%

\end{document}